\documentclass[amsmath,amssymb,superscriptaddress,a4paper,aip,apl,preprint]{revtex4-1}

\usepackage{graphicx} 
\usepackage{amsmath}
\usepackage{amssymb}
\usepackage{amsfonts}
\usepackage{lmodern}
\usepackage{textcomp}
\usepackage[T1]{fontenc}
\usepackage{gensymb} 
\usepackage[seperr, load={}]{siunitx}

\newcommand{\T}{\mathcal{T}}

\begin{document}

\title{Nongalvanic thermometry for ultracold two-dimensional electron domains}

\author{S. Gasparinetti}
\email{simone.gasparinetti@aalto.fi}
\affiliation{Low Temperature Laboratory, Aalto University, P.O. Box 15100, FI-00076 Aalto, Finland}
\author{M. J. Mart\'inez-P\'erez}
\affiliation{NEST Istituto Nanoscienze-CNR and Scuola Normale Superiore, I-56127 Pisa, Italy}
\author{S. de Franceschi}
\affiliation{SPSMS, CEA-INAC/UJF-Grenoble 1, 17 Rue des Martyrs, F-38054 Grenoble Cedex 9, France}
\author{J. P. Pekola}
\affiliation{Low Temperature Laboratory, Aalto University, P.O. Box 15100, FI-00076 Aalto, Finland}
\author{F. Giazotto}
\affiliation{NEST Istituto Nanoscienze-CNR and Scuola Normale Superiore, I-56127 Pisa, Italy}


\begin{abstract}
Measuring the temperature of a two-dimensional electron gas at temperatures of a few mK is a challenging issue, which standard thermometry schemes may fail to tackle. We propose and analyze a nongalvanic thermometer, based on a quantum point contact and quantum dot, which delivers virtually no power to the electron system to be measured.
\end{abstract}

\maketitle

The availability of high-mobility two-dimensional electron gases (2DEGs), combined with the ability to cool them down to low temperatures, has led to the discovery of outstanding physical phenomena, such as the quantum Hall effect \cite{Ando1982}.
Refrigeration schemes are currently under investigation to cool the 2DEG below the conventional operating temperature of a dilution fridge (around 20 mK), down to 1 mK or below \cite{Pan1999, Zumbuhl2011}. This achievement would open the way to a range of experiments of fundamental relevance and to a number of applications:
electron interferometry \cite{Ji2003}, novel correlated phases \cite{Simon2007} and exotic effects \cite{Potok2007}, charge pumping \cite{Giblin2010}, quantum computing \cite{Hanson2007,Nayak2008}, and so on.

As the temperature of electrons gets down to the mK range and below, finding a proper way to measure it in a non-invasive way becomes a critical issue. With the coupling between electrons and phonons becoming weaker and weaker, the power load that a micrometer-sized electron domain can sustain without overheating shrinks down to a few aW or less. In this regime, detection schemes based on transport measurements, such as the ``conventional''  quantum dot thermometer (QDT), become impractical as they inject high-energy quasiparticles which heat the system up, when not bringing it out of thermal equilibrium.

In this Letter, we propose nongalvanic thermometry for 2DEGs. We start with a quick review of the QDT. Then, we introduce its nongalvanic counterpart, whose building blocks are a quantum dot (QD) and a quantum point contact (QPC). This device delivers virtually no power to the electron domain to be measured. We model its operation with standard theory and analyze its performance by choosing realistic parameters. Finally, we discuss the problem of measurement back-action.

\begin{figure}
\center
\includegraphics{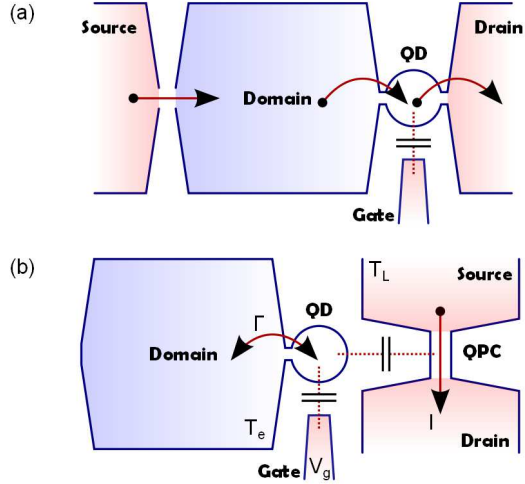}
\caption{(Color online) Galvanic (a) versus nongalvanic (b) QDT. In (a), temperature is determined by the linewidth of Coulomb-blockade peaks, obtained from a transport measurement. In (b), from the average occupation of the dot, read out in a nongalvanic fashion by a QPC placed nearby.}
\label{fig:1}
\end{figure}

\smallskip

An implementation of the QDT is shown in Fig. 1(a). The QD, typically defined by split-gate confinement, is connected by tunnel barriers to two distinct 2DEG regions, one of which is the electron domain to be measured. At zero bias, every time a resonant level of the dot crosses the Fermi energy of the leads, the conductance displays a Coulomb-blockade peak. \cite{Beenakker1991} If the two leads share the same temperature, the latter is simply determined from the peak linewidth, to which it is proportional. On the other hand, when the temperature of the source and drain leads are different, one can still detect the two temperatures independently by applying a voltage bias much greater than the thermal energy of the hotter lead, or even with a single zero-bias measurement, provided the temperature difference is large enough \cite{Gasparinetti2011}.

Based on a transport measurement, this scheme unavoidably brings in dissipation. Of the total power dissipated during the operation of the thermometer, let us estimate the fraction $\dot{Q}_R$ that goes to the domain. This is associated to the tunneling of hot quasiparticles, contributing a heat flow $\dot{Q}_{R} = \Gamma E f(E)$, where $\Gamma$ is the tunneling rate for the resonant level of the dot, $E$ its energy (with respect to the Fermi energy of the domain), and $f$ the electron distribution function in the domain. We shall assume that a quasiequilibrium regime \cite{Giazotto2006} holds, so that $f(E)=[1+\exp(E/k_B T_e)]^{-1}$, $T_e$ being the temperature of the domain.

To perform the readout, we must vary $E$ at least in the range of $-3k_BT_e$ and $3k_BT_e$. Averaging over such a sweep, we obtain $\langle \dot{Q}_{R} \rangle  \approx 0.55 \Gamma k_B T_e$. Now, a lower bound for $\Gamma$ comes from the need for adequate signal-to-noise ratio, the current at resonance being of the order of $e\Gamma$. If we set $1$ pA as a minimum value, we get $\Gamma>\SI{10}{MHz}$. On the other hand, Coulomb-blockade thermometry requires thermal broadening of the peak to dominate above intrinsic (Lorentzian) broadening. This condition, which must hold regardless of dissipation, reads $h\Gamma \ll k_BT_e$; for $T_e=10$ mK, it gives $\Gamma \ll $ 200 MHz.

In the following, we will assume $\Gamma=\SI{10}{MHz}$, which according to our estimate corresponds to $\dot{Q}_{R}/T_e \approx 80 $ aW/K. This figure must be compared to the cooling power provided by all relevant heat-relaxation channels. For definiteness, let us take as the electron domain a
portion of a GaAs/AlGaAs 2DEG of representative density and mobility.
At subkelvin temperatures, the heat flow from electrons into phonons is given (for GaAs-based 2DEGs) by the expression $\dot{Q}_{e-ph} = \Sigma A (T_e^5-T_{ph}^5)$ \cite{Price1982}, where $T_{ph}$ is the temperature of the phonon bath, $A$ the area of the domain, and $\Sigma$ a constant of the order of \SI{30}{fW \micro m^{-2} K^{-5}}.
\cite{Appleyard1998,Gasparinetti2011}

\begin{figure}
\center
\includegraphics{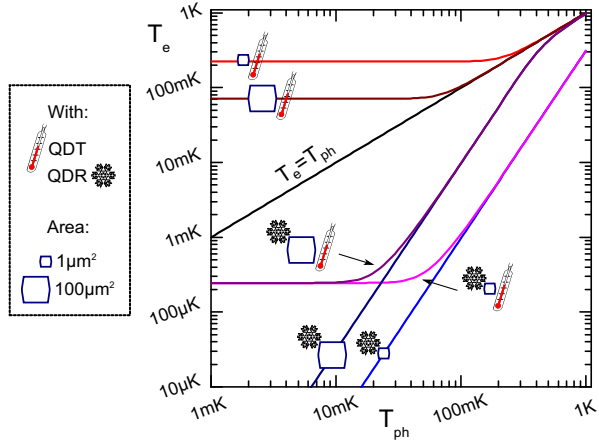}
\caption{(Color online) Steady-state electron temperature $T_e$ versus phonon bath temperature $T_{ph}$, for domains of different areas, in the presence of a QDT, a QDR, or both. Parasitic heat loads on the system are not taken into account (see text).}
\label{fig:powerplot}
\end{figure}

In Figure \ref{fig:powerplot}, we plot the steady-state $T_e$ for 1 and \SI{100}{\micro m^2}-sized domains, versus $T_{ph}$. $T_e$ is determined from a power balance equation of the form $\sum_i \dot{Q}_i[T_e] = 0$, with $Q_i$ denoting the heat flow into the domain due to the $i$th channel.

Each curve refers to a different configuration, to be discussed below.
The straight line marked $T_e=T_{ph}$ is plotted for reference, and stands for the case where no additional heat load is put on the domain. As soon as the QDT is introduced, the situation changes dramatically: $T_e$ follows $T_{ph}$ only down to about 100 mK, below which a saturation occurs. This is due to the coupling between electrons and phonons getting weaker at lower temperatures, a well-known fact which has recently motivated the development of electronic coolers. We take this possibility into account by considering the case where a quantum-dot refrigerator (QDR) \cite{Edwards1993,Edwards1995,Prance2009} is used to cool down the domain, both in the presence and in the absence of the QDT. For simplicity, we assume that the QDR is operated in ideal conditions, so that its cooling power is given by the expression \cite{Edwards1995} $Q_{QDR}=C T_e^2$, with $C \approx \SI{0.31}{pW/K^2}$. Thanks to the QDR, the curves with QDT+QDR now saturate at much lower temperatures, of the order of \SI{1}{mK} or below. Notice that the saturating $T_e$ no longer depends on the domain area; this is because at such low temperatures the competition is between the QDT and the QDR, the phonon bath playing little or no role. 
For simplicity, in the discussion above we have included no other sources of heat besides the QDT. In reality, the electronic temperature is eventually limited by parasitic heat sources, such as radiation from higher-temperature stages and noise in the electrical lines. Likewise, the performance assumed for the QDR must also be taken as an idealization: a recent experiment \cite{Prance2009} pointed out deviations from the ideal behavior already at \SI{110}{mK}, possibly due to nonequilibrium effects.

\smallskip

The nongalvanic device that we propose is shown in Fig.~1(b). As in the QDT discussed above, the strongly nonlinear density of states of a QD is exploited to probe the energy distribution of the domain.
All the difference lies in the way this information is read out: instead of performing a transport measurement across the dot, we measure its average occupation in a nongalvanic fashion with the help of a QPC placed nearby \cite{Field1993,Elzerman2004,DiCarlo2004,note_Olli}.
If the gate sweep is performed adiabatically, the heat flow into the domain is minimal, making the nongalvanic thermometer a candidate device for temperature measurements of ultracold electron domains. In the following, we will describe its operation with a quantitative model.


Let us start from the QD. The latter is preferably operated in the ``quantum'' Coulomb blockade regime, meaning that both its charging energy and orbital level spacing are much greater than the thermal energy. As a result, electron transitions between the dot and the domain involve a single energy level, whose mean occupation number can be written as
\begin{equation} 
\langle n_{dot} \rangle = f(E_0^{QD}- e \alpha V_G)\ ,
\end{equation} 
where $E_0^{QD}$ is a reference energy for the level, and $\alpha$ the lever arm of the gate on the dot.

Our next question is how the change in $\langle n_{dot} \rangle$ affects the current $I$ through the QPC, in the presence of a voltage bias $V_b$. In the Landauer-B\"uttiker formalism \cite{Landauer1988}, $I= \frac{2e}{h} \int_{-\infty}^{\infty} dE \T(E,E^{QPC}) \left[ f(E-e V_b,T_L)-f(E,T_L) \right]$,
where $T_L$ is the temperature of the QPC leads (in general, $T_e \neq T_L$)
and $\T(E,E^{QPC})$ is the energy-dependent transmission coefficient of the QPC. Assuming a single ballistic channel and using a saddle potential \cite{Buttiker1990},
$\T(E,E^{QPC})=\left\{1+\exp \left[ - 2\pi \left( E -E^{QPC} \right) /\hbar \omega_x \right] \right\}^{-1}$,
where $\omega_x$ is a characteristic energy of the confinment and $E^{QPC}$ denotes the bottom of the potential for the one-dimensional electron channel defined by the QPC.
Upon changing $V_g$, the potential landscape at the QPC changes due to the capacitive couplings QPC-QD and QPC-gate. As these couplings are small, we regard them as perturbations and model their effect by a shift of the potential $E^{QPC}$ with respect to a reference value $E_0^{QPC}$. The latter is tuned by the gates defining the constriction, and defines the working point of the QPC.
We shall further denote by $\beta$ the lever arm of the dot on the QPC, and by $\gamma$ that of the gate. In general, we expect $\gamma \ll \beta$. Then we write $E^{QPC}$ as:
\begin{equation} 
E^{QPC} = \beta \frac{e^2 \langle n_{dot} \rangle}{C_\Sigma} - e\gamma V_G -E_0^{QPC}\ .
\end{equation} 
In the limit $eV_b,k_BT_L \ll h\omega_x$,
$\T$ is approximately constant in the range where the electron distributions of the leads vary.
The expression for $I$ then simplifies as $I = \frac{2e^2}{h} \T(0,E^{QPC}) V_b\ .$ Notice that $T_L$ no longer appears in this expression. By contrast, $T_e$ determines $\langle n_{dot} \rangle$, which affects $E^{QPC}$ and hence $\T$.

\begin{figure}
\center
\includegraphics{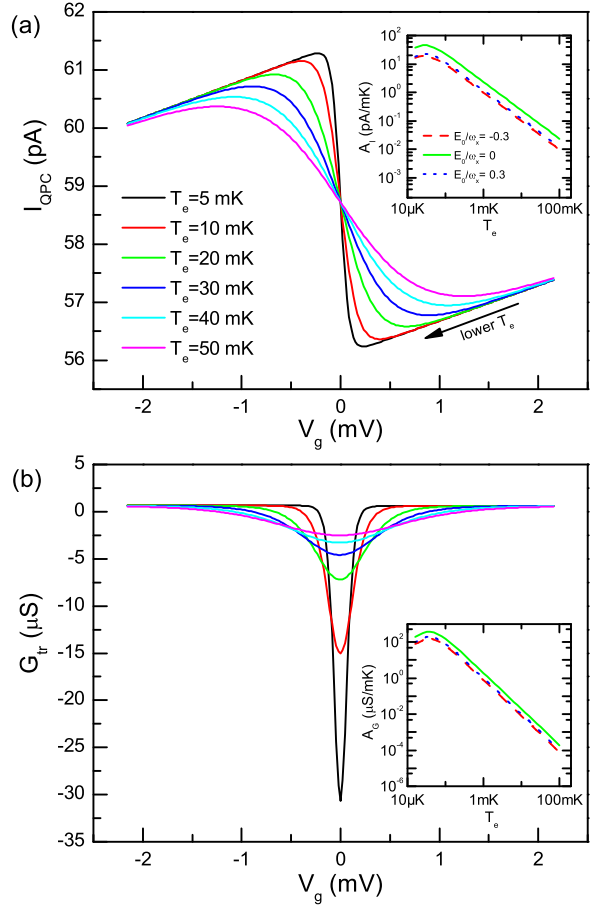}
\caption{(Color online)
(a) QPC current $I$ versus gate voltage $V_g$ for different values of the domain temperature $T_e$; a steeper sawtooth corresponds to a lower $T_e$. Inset: Current gain $A_I$ versus $T_e$ for three different QPC working points. (b) Transconductance $G_{tr}$ versus $V_G$ for the same set of temperatures as in (a); a sharper peak corresponds to a lower $T_e$. Inset: Transconductance gain $A_{G}$ versus $T_e$ [same working points as in (a)].
Parameters: 
$T_{L}= \SI{20}{mK}$,
$\omega_x=\SI{1}{meV}$,
$E_C=\SI{2}{K}$,
$\beta=0.1$, 
$\alpha=0.01$, 
$\gamma=0.002$. 
In the main panels, $E_0/\omega_x=-0.3$. In the Insets, the gains are evaluated at optimal $V_g$ points.
For $A_I$, we take into account \SI{1}{\micro V} fluctuations of $V_g$. For $A_G$, the curves are those expected for a lock-in measurement with \SI{1}{\micro V} signal amplitude.}
\label{fig:performance}
\end{figure}

In Fig.~\ref{fig:performance}(a) we plot $I$ versus $V_g$ for different $T_e$.
As $V_g$ is made more negative, $I$ steadily decreases due to the spurious coupling between the gate and the QPC. Yet as the resonant level crosses the Fermi energy of the domain from above, $\langle n_{dot}\rangle$ sharply decreases by one, leading to a step-like increase in $I$. This gives rise to a sawtooth pattern at zero temperature, which gets progressively smeared as $T_e$ is increased.

Besides $I$, a relevant quantity for thermometry is the gate-to-QPC transconductance $G_{tr}=d I/ d V_g$, which can be directly measured  using a lock-in amplifier. By direct calculation, we find:
\begin{equation}
G_{tr}=\frac{2e^2}{h} e V_b \frac{d\T}{dE^{QPC}}
\left( \gamma + \alpha \beta \frac{e^2}{C_\Sigma} \frac{df}{dE}\right) \ .
\end{equation}
As a function of $V_g$, a series of dips appear on top of a positive baseline [see Fig.~3(b)]. The dips are proportional to the derivative of the Fermi function, and their FWHM $\Delta V_g$ to the domain temperature $T_e$.
Explicitly:
\begin{equation}
T_e=\frac{e\alpha}{2\log(3+2\sqrt{2})k_B} \Delta V_g\ .
\end{equation}
The constant relating $\Delta V_g$ to $T_e$ is a simple combination of fundamental constants and the lever arm $\alpha$, which can be determined experimentally from a measurement of the QD charging energy and the cross-capacitance between the gate and the QD. 
This fact makes of the nongalvanic QDT a primary thermometer, i.e. a thermometer that can measure absolute temperatures without relying on other thermometers (e.g. for calibration). \cite{Giazotto2006}

We conclude this discussion by giving a figure of merit for each measurement mode. If we choose to measure $I$, such a figure may well be the current gain $A_I=\delta I/ \delta T_e$, the ratio being taken at the gate position $V_g^{opt}$ that maximizes it. In the Inset of Fig.~3(a), $A_I$ is plotted versus $T_e$ over a broad range of temperatures, and for different QPC working points. The maximum gain is obtained by choosing $E_0^{QPC}=0$, which corresponds to $\T = \frac{1}{2}$. At \SI{100}{\micro K}, it can exceed 10pA/mK.
Since $A_I$ scales as the inverse of $T_e$, the lower the temperature, the higher the gain. Yet, the sharpness of the sawtooth also increases at lower temperature, so that the measurement becomes more and more sensitive to the dot potential. Fluctuations of $V_g$ of the order of \SI{1}{\micro V}, included in the model, are responsible for the bending of the curves below \SI{50}{\micro K}.
As for $G_{tr}$, we can proceed in the same way and define a transconductance gain $A_{G}=\delta G_{tr}/ \delta T_e$. $A_{G}$ is plotted in the Inset of Fig.~3(b). Similarly to $A_I$, $A_{G}$ is also maximized when $\T = \frac{1}{2}$. At \SI{100}{\micro K}, $A_G \approx \SI{100}{\micro S/mK}$. The dependence on $T_e$ is the same as for $A_I$. At very low $T_e$, $A_G$ is eventually limited by the amplitude of the lock-in modulation applied to $V_g$.

\smallskip

So far, we have implicitly assumed that the state of the QD is not influenced by our readout procedure; that is, we have neglected any measurement backaction. In the following, we shall take it into account and show that its effects are indeed negligible in a suitable range of parameters.
In doing so, we are led to consider two different mechanisms: current fluctuations through the QPC (that is, shot noise) \cite{Blanter2000,Onac2006,Gustavsson2007,Gustavsson2008}, and charge fluctuations in the QPC \cite{Pedersen1998,Young2010}. The nature of these two is very different. In particular, the way current fluctuations couple to the dot depends on the specific setup.
By contrast, the backaction due to charge fluctuations is fundamentally unavoidable. Indeed, it is related to the Heisenberg backaction of the detector (QPC) on the quantum system whose state we are measuring (QD) \cite{Young2010}.

We shall describe both mechanisms using the theory of photon-assisted tunneling (PAT) \cite{Ingold1992}. Let $S_{V}(\omega)$ be the spectrum of voltage fluctuations on the dot; the probability of PAT with energy $E$ is then
$P(E)=\frac{1}{h}\int_{-\infty}^{\infty} \exp\left[J(t)+i\frac{E}{\hbar}t \right]dt$,
where the phase-phase correlation function $J(t)$ is related to $S_{V}(\omega)$ by
$J(t)=\frac{2\pi}{\hbar R_K}
\int_{-\infty}^{\infty} \frac{S_{V} (\omega)}{\omega^2} \left(e^{-i\omega t}-1\right) d\omega$.
The modified $\langle n_{dot} \rangle$, accounting for PAT, is given by:
\begin{equation}
\langle n_{dot} \rangle = \int_{-\infty}^{\infty} f(E-E_0^{QD}- e \alpha V_G) P(E) dE \ ,
\end{equation}
which is a convolution of the distribution function of the domain with the $P(E)$ function. Even in the presence of PAT, our previous analysis is still correct provided $P(E)$ is cutoff at some energy $\bar{E} \ll k_BT$, for in that case we can approximate $P(E) \approx \delta(E)$, and recover the unperturbed result.

Let us consider current noise first. Given its spectral density $S_{I}(\omega)$, the spectrum of voltage fluctuations in the dot is obtained by $S_V(\omega) = |Z(\omega)|^2 S_I(\omega)$, where we have introduced a transimpedance $Z$ as in Ref.~\onlinecite{Aguado2000}. As a first approximation, we may write $Z(\omega) \approx Z(0) = \tau R_S\ ,$ where $R_S$ is the resistance of the QPC leads and $\tau$ a lever arm describing the asymmetric coupling between QD and QPC leads.
The behavior of $P(E)$ at finite energies is then given by $P(E)=\frac{2\pi}{R_K} \frac{Z^2 S_I(E/\hbar)}{E^2}$,
where $R_K=h/2e^2$ is the resistance quantum.
Taking normalization into account, we find that the energy spread of $P(E)$ is of the order of $\bar{E}=Z^2S_I/R_K$. Now, shot noise in the QPC has the spectrum \cite{Blanter2000} $S_I=(eV_b/R_K) \T(1-\T)$. If we take $R_S=0.1 R_K$, $\tau=0.1$, $\T=1/2$ and $V_b=\SI{2.5}{\micro V}$ (so that $I=\SI{100}{pA}$), we get $\bar{E}/k_B=\SI{3}{\micro K}$. As revealed by this analysis, the backaction due to current noise can be made negligible by a combination of low-resistance leads and small $V_b$.

Let us now turn to charge noise. 
The spectrum of charge flucutations on the dot, induced by the QPC, is given to the first order in $V_b$ and $\omega$ by the expression $S_{Q}(\omega) = 2 C_{\mu}^2 R_v eV_b$, where $R_v$ is the nonequilibrium charge relaxation resistance defined in Ref.~\onlinecite{Pedersen1998},
and $C_{\mu}$ the electrochemical capacitance of the QPC ``to'' the dot. 
Charge fluctuations are related to voltage fluctuations by the total capacitance of the dot: $S_{V}=(1/C_\Sigma)^2 S_{Q}$, so that
$S_{V}=2 (C_\mu/C_\Sigma)^2 eV_b R_v\ .$ As for current noise, we have $P(E)=\frac{2\pi}{R_K} \frac{S_{V}(E/\hbar)}{E^2}$.
The energy spread for this $P(E)$ is given by $\bar{E}=(C_\mu/C_\Sigma)^2 (R_v/ R_K) eV_b$. We estimate its magnitude by taking $C_\mu/C_\Sigma=0.02$, $R_v=0.1R_K$, $V_b=\SI{2.5}{\micro V}$. We get $\bar{E}/k_B\approx\SI{1}{\micro K}$, implying that we can safely neglect charge noise down to very low temperatures. This primarily stems from the ratio $C_\mu/C_\Sigma$ being very small, as typical for split-gate-defined nanostructures. In addition, the same prescription as for current noise must be applied to $V_b$.

\smallskip

In conclusion, we have addressed the problem of measuring the temperature of 2DEG microdomains cooled down to the base temperature of state-of-art dilution refrigerators, and possibly below. Already at 100 mK, conventional schemes based on transport are inadequate, due to overheating. We have argued that nongalvanic thermometry may overcome this limitation. Our results suggest that a nongalvanic thermometer such as that considered may be conveniently employed at temperatures ranging from tens of mK down to tens of \SI{}{\micro K}.

\smallskip

We would like to thank R.~Aguado, F. Portier and O.-P.~Saira for useful discussions. This work was supported by the European Community FP7 project No.~228464 `Microkelvin' and the Finnish National Graduate School in Nanoscience. S.D.F acknowledges support from the ERC Starting Grant program.


\begin{thebibliography}{30}%
\makeatletter
\providecommand \@ifxundefined [1]{%
 \@ifx{#1\undefined}
}%
\providecommand \@ifnum [1]{%
 \ifnum #1\expandafter \@firstoftwo
 \else \expandafter \@secondoftwo
 \fi
}%
\providecommand \@ifx [1]{%
 \ifx #1\expandafter \@firstoftwo
 \else \expandafter \@secondoftwo
 \fi
}%
\providecommand \natexlab [1]{#1}%
\providecommand \enquote  [1]{``#1''}%
\providecommand \bibnamefont  [1]{#1}%
\providecommand \bibfnamefont [1]{#1}%
\providecommand \citenamefont [1]{#1}%
\providecommand \href@noop [0]{\@secondoftwo}%
\providecommand \href [0]{\begingroup \@sanitize@url \@href}%
\providecommand \@href[1]{\@@startlink{#1}\@@href}%
\providecommand \@@href[1]{\endgroup#1\@@endlink}%
\providecommand \@sanitize@url [0]{\catcode `\\12\catcode `\$12\catcode
  `\&12\catcode `\#12\catcode `\^12\catcode `\_12\catcode `\%12\relax}%
\providecommand \@@startlink[1]{}%
\providecommand \@@endlink[0]{}%
\providecommand \url  [0]{\begingroup\@sanitize@url \@url }%
\providecommand \@url [1]{\endgroup\@href {#1}{\urlprefix }}%
\providecommand \urlprefix  [0]{URL }%
\providecommand \Eprint [0]{\href }%
\@ifxundefined \urlstyle {%
  \providecommand \doi  [0]{\begingroup \@sanitize@url \@doi}%
  \providecommand \@doi [1]{\endgroup \@@startlink {\doibase
  #1}doi:\discretionary {}{}{}#1\@@endlink }%
}{%
  \providecommand \doi  [0]{doi:\discretionary{}{}{}\begingroup
  \urlstyle{rm}\Url }%
}%
\providecommand \doibase [0]{http://dx.doi.org/}%
\providecommand \Doi [0]{\begingroup \@sanitize@url \@Doi }%
\providecommand \@Doi  [1]{\endgroup\@@startlink{\doibase#1}\@@Doi}%
\providecommand \@@Doi [1]{#1\@@endlink}%
\providecommand \selectlanguage [0]{\@gobble}%
\providecommand \bibinfo  [0]{\@secondoftwo}%
\providecommand \bibfield  [0]{\@secondoftwo}%
\providecommand \translation [1]{[#1]}%
\providecommand \BibitemOpen [0]{}%
\providecommand \bibitemStop [0]{}%
\providecommand \bibitemNoStop [0]{.\EOS\space}%
\providecommand \EOS [0]{\spacefactor3000\relax}%
\providecommand \BibitemShut  [1]{\csname bibitem#1\endcsname}%
\bibitem [{\citenamefont {Ando}\ \emph {et~al.}(1982)\citenamefont {Ando},
  \citenamefont {Fowler},\ and\ \citenamefont {Stern}}]{Ando1982}%
  \BibitemOpen
  \bibfield  {author} {\bibinfo {author} {\bibfnamefont {T.}~\bibnamefont
  {Ando}}, \bibinfo {author} {\bibfnamefont {B.}~\bibnamefont {Fowler}}, \ and\
  \bibinfo {author} {\bibfnamefont {F.}~\bibnamefont {Stern}},\ }\href@noop {}
  {\bibfield  {journal} {\bibinfo  {journal} {Rev. Mod. Phys.},\ }\textbf
  {\bibinfo {volume} {54}},\ \bibinfo {pages} {437} (\bibinfo {year}
  {1982})}\BibitemShut {NoStop}%
\bibitem [{\citenamefont {Pan}\ \emph {et~al.}(1999)\citenamefont {Pan},
  \citenamefont {Xia}, \citenamefont {Shvarts}, \citenamefont {Adams},
  \citenamefont {Stormer}, \citenamefont {Tsui}, \citenamefont {Pfeiffer},
  \citenamefont {Baldwin},\ and\ \citenamefont {West}}]{Pan1999}%
  \BibitemOpen
  \bibfield  {author} {\bibinfo {author} {\bibfnamefont {W.}~\bibnamefont
  {Pan}}, \bibinfo {author} {\bibfnamefont {J.}~\bibnamefont {Xia}}, \bibinfo
  {author} {\bibfnamefont {V.}~\bibnamefont {Shvarts}}, \bibinfo {author}
  {\bibfnamefont {D.~E.}\ \bibnamefont {Adams}}, \bibinfo {author}
  {\bibfnamefont {H.~L.}\ \bibnamefont {Stormer}}, \bibinfo {author}
  {\bibfnamefont {D.~C.}\ \bibnamefont {Tsui}}, \bibinfo {author}
  {\bibfnamefont {L.~N.}\ \bibnamefont {Pfeiffer}}, \bibinfo {author}
  {\bibfnamefont {K.~W.}\ \bibnamefont {Baldwin}}, \ and\ \bibinfo {author}
  {\bibfnamefont {K.~W.}\ \bibnamefont {West}},\ }\href@noop {} {\bibfield
  {journal} {\bibinfo  {journal} {Phys. Rev. Lett.},\ }\textbf {\bibinfo
  {volume} {83}},\ \bibinfo {pages} {3530} (\bibinfo {year}
  {1999})}\BibitemShut {NoStop}%
\bibitem [{\citenamefont {Clark}\ \emph {et~al.}(2011)\citenamefont {Clark},
  \citenamefont {Schwarzw\"{a}lder~K}, \citenamefont {Bandi}, \citenamefont
  {Maradan},\ and\ \citenamefont {Zumb\"{u}hl}}]{Zumbuhl2011}%
  \BibitemOpen
  \bibfield  {author} {\bibinfo {author} {\bibfnamefont {A.~C.}\ \bibnamefont
  {Clark}}, \bibinfo {author} {\bibfnamefont {K.}~\bibnamefont
  {Schwarzw\"{a}lder~K}}, \bibinfo {author} {\bibfnamefont {T.}~\bibnamefont
  {Bandi}}, \bibinfo {author} {\bibfnamefont {D.}~\bibnamefont {Maradan}}, \
  and\ \bibinfo {author} {\bibfnamefont {D.~M.}\ \bibnamefont {Zumb\"{u}hl}},\
  }\href@noop {} {} (\bibinfo {year} {2011}),\ \Eprint
  {http://arxiv.org/abs/1111.1972} {arXiv:1111.1972} \BibitemShut {NoStop}%
\bibitem [{\citenamefont {Ji}\ \emph {et~al.}(2003)\citenamefont {Ji},
  \citenamefont {Chung}, \citenamefont {Sprinzak}, \citenamefont {Heiblum},\
  and\ \citenamefont {Mahalu}}]{Ji2003}%
  \BibitemOpen
  \bibfield  {author} {\bibinfo {author} {\bibfnamefont {Y.}~\bibnamefont
  {Ji}}, \bibinfo {author} {\bibfnamefont {Y.}~\bibnamefont {Chung}}, \bibinfo
  {author} {\bibfnamefont {D.}~\bibnamefont {Sprinzak}}, \bibinfo {author}
  {\bibfnamefont {M.}~\bibnamefont {Heiblum}}, \ and\ \bibinfo {author}
  {\bibfnamefont {D.}~\bibnamefont {Mahalu}},\ }\href@noop {} {\bibfield
  {journal} {\bibinfo  {journal} {Nature},\ }\textbf {\bibinfo {volume}
  {422}},\ \bibinfo {pages} {415} (\bibinfo {year} {2003})}\BibitemShut
  {NoStop}%
\bibitem [{\citenamefont {Simon}\ and\ \citenamefont {Loss}(2007)}]{Simon2007}%
  \BibitemOpen
  \bibfield  {author} {\bibinfo {author} {\bibfnamefont {P.}~\bibnamefont
  {Simon}}\ and\ \bibinfo {author} {\bibfnamefont {D.}~\bibnamefont {Loss}},\
  }\href@noop {} {\bibfield  {journal} {\bibinfo  {journal} {Phys. Rev.
  Lett.},\ }\textbf {\bibinfo {volume} {98}},\ \bibinfo {pages} {156401}
  (\bibinfo {year} {2007})}\BibitemShut {NoStop}%
\bibitem [{\citenamefont {Potok}\ \emph {et~al.}(2007)\citenamefont {Potok},
  \citenamefont {Rau}, \citenamefont {Shtrikman}, \citenamefont {Oreg},\ and\
  \citenamefont {Goldhaber-Gordon}}]{Potok2007}%
  \BibitemOpen
  \bibfield  {author} {\bibinfo {author} {\bibfnamefont {R.~M.}\ \bibnamefont
  {Potok}}, \bibinfo {author} {\bibfnamefont {I.~G.}\ \bibnamefont {Rau}},
  \bibinfo {author} {\bibfnamefont {H.}~\bibnamefont {Shtrikman}}, \bibinfo
  {author} {\bibfnamefont {Y.}~\bibnamefont {Oreg}}, \ and\ \bibinfo {author}
  {\bibfnamefont {D.}~\bibnamefont {Goldhaber-Gordon}},\ }\href@noop {}
  {\bibfield  {journal} {\bibinfo  {journal} {Nature},\ }\textbf {\bibinfo
  {volume} {446}},\ \bibinfo {pages} {167} (\bibinfo {year}
  {2007})}\BibitemShut {NoStop}%
\bibitem [{\citenamefont {Giblin}\ \emph {et~al.}(2010)\citenamefont {Giblin},
  \citenamefont {Wright}, \citenamefont {Fletcher}, \citenamefont {Kataoka},
  \citenamefont {Pepper}, \citenamefont {Janssen}, \citenamefont {Ritchie},
  \citenamefont {Nicoll}, \citenamefont {Anderson},\ and\ \citenamefont
  {Jones}}]{Giblin2010}%
  \BibitemOpen
  \bibfield  {author} {\bibinfo {author} {\bibfnamefont {S.~P.}\ \bibnamefont
  {Giblin}}, \bibinfo {author} {\bibfnamefont {S.~J.}\ \bibnamefont {Wright}},
  \bibinfo {author} {\bibfnamefont {J.~D.}\ \bibnamefont {Fletcher}}, \bibinfo
  {author} {\bibfnamefont {M.}~\bibnamefont {Kataoka}}, \bibinfo {author}
  {\bibfnamefont {M.}~\bibnamefont {Pepper}}, \bibinfo {author} {\bibfnamefont
  {T.~J. B.~M.}\ \bibnamefont {Janssen}}, \bibinfo {author} {\bibfnamefont
  {D.~A.}\ \bibnamefont {Ritchie}}, \bibinfo {author} {\bibfnamefont {C.~A.}\
  \bibnamefont {Nicoll}}, \bibinfo {author} {\bibfnamefont {D.}~\bibnamefont
  {Anderson}}, \ and\ \bibinfo {author} {\bibfnamefont {G.~A.~C.}\ \bibnamefont
  {Jones}},\ }\href@noop {} {\bibfield  {journal} {\bibinfo  {journal} {New
  Journ. Phys.},\ }\textbf {\bibinfo {volume} {12}},\ \bibinfo {pages} {073013}
  (\bibinfo {year} {2010})}\BibitemShut {NoStop}%
\bibitem [{\citenamefont {Hanson}\ \emph {et~al.}(2007)\citenamefont {Hanson},
  \citenamefont {Kouwenhoven}, \citenamefont {Petta}, \citenamefont {Tarucha},\
  and\ \citenamefont {Vandersypen}}]{Hanson2007}%
  \BibitemOpen
  \bibfield  {author} {\bibinfo {author} {\bibfnamefont {R.}~\bibnamefont
  {Hanson}}, \bibinfo {author} {\bibfnamefont {L.~P.}\ \bibnamefont
  {Kouwenhoven}}, \bibinfo {author} {\bibfnamefont {J.}~\bibnamefont {Petta}},
  \bibinfo {author} {\bibfnamefont {S.}~\bibnamefont {Tarucha}}, \ and\
  \bibinfo {author} {\bibfnamefont {L.~M.~K.}\ \bibnamefont {Vandersypen}},\
  }\href@noop {} {\bibfield  {journal} {\bibinfo  {journal} {Rev. Mod. Phys.},\
  }\textbf {\bibinfo {volume} {79}},\ \bibinfo {pages} {1217} (\bibinfo {year}
  {2007})}\BibitemShut {NoStop}%
\bibitem [{\citenamefont {Nayak}\ \emph {et~al.}(2008)\citenamefont {Nayak},
  \citenamefont {Stern}, \citenamefont {Freedman},\ and\ \citenamefont {{Das
  Sarma}}}]{Nayak2008}%
  \BibitemOpen
  \bibfield  {author} {\bibinfo {author} {\bibfnamefont {C.}~\bibnamefont
  {Nayak}}, \bibinfo {author} {\bibfnamefont {A.}~\bibnamefont {Stern}},
  \bibinfo {author} {\bibfnamefont {M.}~\bibnamefont {Freedman}}, \ and\
  \bibinfo {author} {\bibfnamefont {S.}~\bibnamefont {{Das Sarma}}},\
  }\href@noop {} {\bibfield  {journal} {\bibinfo  {journal} {Rev. Mod. Phys.},\
  }\textbf {\bibinfo {volume} {80}},\ \bibinfo {pages} {1083} (\bibinfo {year}
  {2008})}\BibitemShut {NoStop}%
\bibitem [{\citenamefont {Beenakker}(1991)}]{Beenakker1991}%
  \BibitemOpen
  \bibfield  {author} {\bibinfo {author} {\bibfnamefont {C.~W.~J.}\
  \bibnamefont {Beenakker}},\ }\href@noop {} {\bibfield  {journal} {\bibinfo
  {journal} {Phys. Rev. B},\ }\textbf {\bibinfo {volume} {44}},\ \bibinfo
  {pages} {1646} (\bibinfo {year} {1991})}\BibitemShut {NoStop}%
\bibitem [{\citenamefont {Gasparinetti}\ \emph {et~al.}(2011)\citenamefont
  {Gasparinetti}, \citenamefont {Deon}, \citenamefont {Biasiol}, \citenamefont
  {Sorba}, \citenamefont {Beltram},\ and\ \citenamefont
  {Giazotto}}]{Gasparinetti2011}%
  \BibitemOpen
  \bibfield  {author} {\bibinfo {author} {\bibfnamefont {S.}~\bibnamefont
  {Gasparinetti}}, \bibinfo {author} {\bibfnamefont {F.}~\bibnamefont {Deon}},
  \bibinfo {author} {\bibfnamefont {G.}~\bibnamefont {Biasiol}}, \bibinfo
  {author} {\bibfnamefont {L.}~\bibnamefont {Sorba}}, \bibinfo {author}
  {\bibfnamefont {F.}~\bibnamefont {Beltram}}, \ and\ \bibinfo {author}
  {\bibfnamefont {F.}~\bibnamefont {Giazotto}},\ }\href@noop {} {\bibfield
  {journal} {\bibinfo  {journal} {Phys. Rev. B},\ }\textbf {\bibinfo {volume}
  {83}},\ \bibinfo {pages} {201306(R)} (\bibinfo {year} {2011})}\BibitemShut
  {NoStop}%
\bibitem [{\citenamefont {Giazotto}\ \emph {et~al.}(2006)\citenamefont
  {Giazotto}, \citenamefont {Heikkil\"{a}}, \citenamefont {Luukanen},
  \citenamefont {Savin},\ and\ \citenamefont {Pekola}}]{Giazotto2006}%
  \BibitemOpen
  \bibfield  {author} {\bibinfo {author} {\bibfnamefont {F.}~\bibnamefont
  {Giazotto}}, \bibinfo {author} {\bibfnamefont {T.~T.}\ \bibnamefont
  {Heikkil\"{a}}}, \bibinfo {author} {\bibfnamefont {A.}~\bibnamefont
  {Luukanen}}, \bibinfo {author} {\bibfnamefont {A.~M.}\ \bibnamefont {Savin}},
  \ and\ \bibinfo {author} {\bibfnamefont {J.~P.}\ \bibnamefont {Pekola}},\
  }\href@noop {} {\bibfield  {journal} {\bibinfo  {journal} {Rev. Mod. Phys.},\
  }\textbf {\bibinfo {volume} {78}},\ \bibinfo {pages} {217} (\bibinfo {year}
  {2006})}\BibitemShut {NoStop}%
\bibitem [{\citenamefont {Price}(1982)}]{Price1982}%
  \BibitemOpen
  \bibfield  {author} {\bibinfo {author} {\bibfnamefont {P.}~\bibnamefont
  {Price}},\ }\href@noop {} {\bibfield  {journal} {\bibinfo  {journal} {J.
  Appl. Phys.},\ }\textbf {\bibinfo {volume} {53}},\ \bibinfo {pages} {6863}
  (\bibinfo {year} {1982})}\BibitemShut {NoStop}%
\bibitem [{\citenamefont {Appleyard}\ \emph {et~al.}(1998)\citenamefont
  {Appleyard}, \citenamefont {Nicholls}, \citenamefont {Simmons}, \citenamefont
  {Tribe},\ and\ \citenamefont {Pepper}}]{Appleyard1998}%
  \BibitemOpen
  \bibfield  {author} {\bibinfo {author} {\bibfnamefont {N.~J.}\ \bibnamefont
  {Appleyard}}, \bibinfo {author} {\bibfnamefont {J.~T.}\ \bibnamefont
  {Nicholls}}, \bibinfo {author} {\bibfnamefont {M.~Y.}\ \bibnamefont
  {Simmons}}, \bibinfo {author} {\bibfnamefont {W.~R.}\ \bibnamefont {Tribe}},
  \ and\ \bibinfo {author} {\bibfnamefont {M.}~\bibnamefont {Pepper}},\
  }\href@noop {} {\bibfield  {journal} {\bibinfo  {journal} {Phys. Rev.
  Lett.},\ }\textbf {\bibinfo {volume} {81}},\ \bibinfo {pages} {3491}
  (\bibinfo {year} {1998})}\BibitemShut {NoStop}%
\bibitem [{\citenamefont {Edwards}\ \emph {et~al.}(1993)\citenamefont
  {Edwards}, \citenamefont {Niu},\ and\ \citenamefont
  {de~Lozanne}}]{Edwards1993}%
  \BibitemOpen
  \bibfield  {author} {\bibinfo {author} {\bibfnamefont {H.~L.}\ \bibnamefont
  {Edwards}}, \bibinfo {author} {\bibfnamefont {Q.}~\bibnamefont {Niu}}, \ and\
  \bibinfo {author} {\bibfnamefont {A.~L.}\ \bibnamefont {de~Lozanne}},\
  }\href@noop {} {\bibfield  {journal} {\bibinfo  {journal} {Appl. Phys.
  Lett.},\ }\textbf {\bibinfo {volume} {63}},\ \bibinfo {pages} {1815}
  (\bibinfo {year} {1993})}\BibitemShut {NoStop}%
\bibitem [{\citenamefont {Edwards}\ \emph {et~al.}(1995)\citenamefont
  {Edwards}, \citenamefont {Niu}, \citenamefont {Georgakis},\ and\
  \citenamefont {de~Lozanne}}]{Edwards1995}%
  \BibitemOpen
  \bibfield  {author} {\bibinfo {author} {\bibfnamefont {H.~L.}\ \bibnamefont
  {Edwards}}, \bibinfo {author} {\bibfnamefont {Q.}~\bibnamefont {Niu}},
  \bibinfo {author} {\bibfnamefont {G.~A.}\ \bibnamefont {Georgakis}}, \ and\
  \bibinfo {author} {\bibfnamefont {A.~L.}\ \bibnamefont {de~Lozanne}},\
  }\href@noop {} {\bibfield  {journal} {\bibinfo  {journal} {Phys. Rev. B},\
  }\textbf {\bibinfo {volume} {52}},\ \bibinfo {pages} {5714} (\bibinfo {year}
  {1995})}\BibitemShut {NoStop}%
\bibitem [{\citenamefont {Prance}\ \emph {et~al.}(2009)\citenamefont {Prance},
  \citenamefont {Smith}, \citenamefont {Griffiths}, \citenamefont {Chorley},
  \citenamefont {Anderson}, \citenamefont {Jones}, \citenamefont {Farrer},\
  and\ \citenamefont {Ritchie}}]{Prance2009}%
  \BibitemOpen
  \bibfield  {author} {\bibinfo {author} {\bibfnamefont {J.~R.}\ \bibnamefont
  {Prance}}, \bibinfo {author} {\bibfnamefont {C.~G.}\ \bibnamefont {Smith}},
  \bibinfo {author} {\bibfnamefont {J.~P.}\ \bibnamefont {Griffiths}}, \bibinfo
  {author} {\bibfnamefont {S.~J.}\ \bibnamefont {Chorley}}, \bibinfo {author}
  {\bibfnamefont {D.}~\bibnamefont {Anderson}}, \bibinfo {author}
  {\bibfnamefont {G.~A.~C.}\ \bibnamefont {Jones}}, \bibinfo {author}
  {\bibfnamefont {I.}~\bibnamefont {Farrer}}, \ and\ \bibinfo {author}
  {\bibfnamefont {D.~A.}\ \bibnamefont {Ritchie}},\ }\href@noop {} {\bibfield
  {journal} {\bibinfo  {journal} {Phys. Rev. Lett.},\ }\textbf {\bibinfo
  {volume} {102}},\ \bibinfo {pages} {146602} (\bibinfo {year}
  {2009})}\BibitemShut {NoStop}%
\bibitem [{\citenamefont {Field}\ \emph {et~al.}(1993)\citenamefont {Field},
  \citenamefont {Smith}, \citenamefont {Pepper}, \citenamefont {Ritchie},
  \citenamefont {Frost}, \citenamefont {Jones},\ and\ \citenamefont
  {Hasko}}]{Field1993}%
  \BibitemOpen
  \bibfield  {author} {\bibinfo {author} {\bibfnamefont {M.}~\bibnamefont
  {Field}}, \bibinfo {author} {\bibfnamefont {C.~G.}\ \bibnamefont {Smith}},
  \bibinfo {author} {\bibfnamefont {M.}~\bibnamefont {Pepper}}, \bibinfo
  {author} {\bibfnamefont {D.~A.}\ \bibnamefont {Ritchie}}, \bibinfo {author}
  {\bibfnamefont {J.~E.~F.}\ \bibnamefont {Frost}}, \bibinfo {author}
  {\bibfnamefont {G.~A.~C.}\ \bibnamefont {Jones}}, \ and\ \bibinfo {author}
  {\bibfnamefont {D.~G.}\ \bibnamefont {Hasko}},\ }\href@noop {} {\bibfield
  {journal} {\bibinfo  {journal} {Phys. Rev. Lett.},\ }\textbf {\bibinfo
  {volume} {70}},\ \bibinfo {pages} {1311} (\bibinfo {year}
  {1993})}\BibitemShut {NoStop}%
\bibitem [{\citenamefont {Elzerman}\ \emph {et~al.}(2004)\citenamefont
  {Elzerman}, \citenamefont {Hanson}, \citenamefont {{Willems van Beveren}},
  \citenamefont {Witkamp}, \citenamefont {Vandersypen},\ and\ \citenamefont
  {Kouwenhoven}}]{Elzerman2004}%
  \BibitemOpen
  \bibfield  {author} {\bibinfo {author} {\bibfnamefont {J.~M.}\ \bibnamefont
  {Elzerman}}, \bibinfo {author} {\bibfnamefont {R.}~\bibnamefont {Hanson}},
  \bibinfo {author} {\bibfnamefont {L.~H.}\ \bibnamefont {{Willems van
  Beveren}}}, \bibinfo {author} {\bibfnamefont {B.}~\bibnamefont {Witkamp}},
  \bibinfo {author} {\bibfnamefont {L.~M.~K.}\ \bibnamefont {Vandersypen}}, \
  and\ \bibinfo {author} {\bibfnamefont {L.~P.}\ \bibnamefont {Kouwenhoven}},\
  }\href@noop {} {\bibfield  {journal} {\bibinfo  {journal} {Nature},\ }\textbf
  {\bibinfo {volume} {430}},\ \bibinfo {pages} {431} (\bibinfo {year}
  {2004})}\BibitemShut {NoStop}%
\bibitem [{\citenamefont {Di Carlo}\ \emph {et~al.}(2004)\citenamefont
  {Di Carlo}, \citenamefont {Lynch}, \citenamefont {{Johnson}},
  \citenamefont {Childress}, \citenamefont {Crockett},\citenamefont {Marcus},\citenamefont {Hanson},\ and\ \citenamefont
  {Gossard}}]{DiCarlo2004}%
  \BibitemOpen
  \bibfield  {author} {\bibinfo {author} {\bibfnamefont {L.}\ \bibnamefont
  {Di Carlo}}, \bibinfo {author} {\bibfnamefont {H.~J.}~\bibnamefont {Lynch}},
  \bibinfo {author} {\bibfnamefont {A.~C.}\ \bibnamefont {{Johnson}}},
  \bibinfo {author} {\bibfnamefont {L.~I.}~\bibnamefont {Childress}},
  \bibinfo {author} {\bibfnamefont {K.}~\bibnamefont {Crockett}},
  \bibinfo {author} {\bibfnamefont {C.~M.}~\bibnamefont {Marcus}},
  \bibinfo {author} {\bibfnamefont {M.~P.}~\bibnamefont {Hanson}}, \
  and\ \bibinfo {author} {\bibfnamefont {A.~C.}\ \bibnamefont {Gossard}},\
  }\href@noop {} {\bibfield  {journal} {\bibinfo  {journal} {Phys. Rev. Lett.},\ }\textbf
  {\bibinfo {volume} {92}},\ \bibinfo {pages} {226801} (\bibinfo {year}
  {2004})}\BibitemShut {NoStop}%
\bibitem [{note_Olli()}]{note_Olli}%
  \BibitemOpen
  \bibinfo {note} {The principle of nongalvanic thermometry may as well be
  applied to metallic systems, with a single-electron transistor playing the
  role of the QPC, as in O.-P. Saira, A. Kemppinen, V. Maisi, and J. Pekola,
  Phys. Rev. B \textbf{85}, 012504 (2012).}\BibitemShut {Stop}%
\bibitem [{\citenamefont {Landauer}(1988)}]{Landauer1988}%
  \BibitemOpen
  \bibfield  {author} {\bibinfo {author} {\bibfnamefont {R.}~\bibnamefont
  {Landauer}},\ }\href@noop {} {\bibfield  {journal} {\bibinfo  {journal} {IBM
  J. Res. Dev.},\ }\textbf {\bibinfo {volume} {32}},\ \bibinfo {pages} {306}
  (\bibinfo {year} {1988})}\BibitemShut {NoStop}%
\bibitem [{\citenamefont {B\"{u}ttiker}(1990)}]{Buttiker1990}%
  \BibitemOpen
  \bibfield  {author} {\bibinfo {author} {\bibfnamefont {M.}~\bibnamefont
  {B\"{u}ttiker}},\ }\href@noop {} {\bibfield  {journal} {\bibinfo  {journal}
  {Phys. Rev. B},\ }\textbf {\bibinfo {volume} {41}},\ \bibinfo {pages} {7906}
  (\bibinfo {year} {1990})}\BibitemShut {NoStop}%
\bibitem [{\citenamefont {Blanter}\ and\ \citenamefont
  {B\"{u}ttiker}(2000)}]{Blanter2000}%
  \BibitemOpen
  \bibfield  {author} {\bibinfo {author} {\bibfnamefont {Y.}~\bibnamefont
  {Blanter}}\ and\ \bibinfo {author} {\bibfnamefont {M.}~\bibnamefont
  {B\"{u}ttiker}},\ }\href@noop {} {\bibfield  {journal} {\bibinfo  {journal}
  {Physics Reports},\ }\textbf {\bibinfo {volume} {336}},\ \bibinfo {pages} {1}
  (\bibinfo {year} {2000})}\BibitemShut {NoStop}%
\bibitem [{\citenamefont {Onac}\ \emph {et~al.}(2006)\citenamefont {Onac},
  \citenamefont {Balestro}, \citenamefont {van Beveren}, \citenamefont
  {Hartmann}, \citenamefont {Nazarov},\ and\ \citenamefont
  {Kouwenhoven}}]{Onac2006}%
  \BibitemOpen
  \bibfield  {author} {\bibinfo {author} {\bibfnamefont {E.}~\bibnamefont
  {Onac}}, \bibinfo {author} {\bibfnamefont {F.}~\bibnamefont {Balestro}},
  \bibinfo {author} {\bibfnamefont {L.~H.~W.}\ \bibnamefont {van Beveren}},
  \bibinfo {author} {\bibfnamefont {U.}~\bibnamefont {Hartmann}}, \bibinfo
  {author} {\bibfnamefont {Y.~V.}\ \bibnamefont {Nazarov}}, \ and\ \bibinfo
  {author} {\bibfnamefont {L.~P.}\ \bibnamefont {Kouwenhoven}},\ }\href@noop {}
  {\bibfield  {journal} {\bibinfo  {journal} {Phys. Rev. Lett.},\ }\textbf
  {\bibinfo {volume} {96}},\ \bibinfo {pages} {176601} (\bibinfo {year}
  {2006})}\BibitemShut {NoStop}%
\bibitem [{\citenamefont {Gustavsson}\ \emph {et~al.}(2007)\citenamefont
  {Gustavsson}, \citenamefont {Studer}, \citenamefont {Leturcq}, \citenamefont
  {Ihn}, \citenamefont {Ensslin}, \citenamefont {Driscoll},\ and\ \citenamefont
  {Gossard}}]{Gustavsson2007}%
  \BibitemOpen
  \bibfield  {author} {\bibinfo {author} {\bibfnamefont {S.}~\bibnamefont
  {Gustavsson}}, \bibinfo {author} {\bibfnamefont {M.}~\bibnamefont {Studer}},
  \bibinfo {author} {\bibfnamefont {R.}~\bibnamefont {Leturcq}}, \bibinfo
  {author} {\bibfnamefont {T.}~\bibnamefont {Ihn}}, \bibinfo {author}
  {\bibfnamefont {K.}~\bibnamefont {Ensslin}}, \bibinfo {author} {\bibfnamefont
  {D.}~\bibnamefont {Driscoll}}, \ and\ \bibinfo {author} {\bibfnamefont
  {A.}~\bibnamefont {Gossard}},\ }\href@noop {} {\bibfield  {journal} {\bibinfo
   {journal} {Phys. Rev. Lett.},\ }\textbf {\bibinfo {volume} {99}},\ \bibinfo
  {pages} {206804} (\bibinfo {year} {2007})}\BibitemShut {NoStop}%
\bibitem [{\citenamefont {Gustavsson}\ \emph {et~al.}(2008)\citenamefont
  {Gustavsson}, \citenamefont {Shorubalko}, \citenamefont {Leturcq},
  \citenamefont {Ihn}, \citenamefont {Ensslin},\ and\ \citenamefont
  {Sch\"{o}n}}]{Gustavsson2008}%
  \BibitemOpen
  \bibfield  {author} {\bibinfo {author} {\bibfnamefont {S.}~\bibnamefont
  {Gustavsson}}, \bibinfo {author} {\bibfnamefont {I.}~\bibnamefont
  {Shorubalko}}, \bibinfo {author} {\bibfnamefont {R.}~\bibnamefont {Leturcq}},
  \bibinfo {author} {\bibfnamefont {T.}~\bibnamefont {Ihn}}, \bibinfo {author}
  {\bibfnamefont {K.}~\bibnamefont {Ensslin}}, \ and\ \bibinfo {author}
  {\bibfnamefont {S.}~\bibnamefont {Sch\"{o}n}},\ }\href@noop {} {\bibfield
  {journal} {\bibinfo  {journal} {Phys. Rev. B},\ }\textbf {\bibinfo {volume}
  {78}},\ \bibinfo {pages} {035324} (\bibinfo {year} {2008})}\BibitemShut
  {NoStop}%
\bibitem [{\citenamefont {Pedersen}\ \emph {et~al.}(1998)\citenamefont
  {Pedersen}, \citenamefont {{Van Langen}},\ and\ \citenamefont
  {B\"{u}ttiker}}]{Pedersen1998}%
  \BibitemOpen
  \bibfield  {author} {\bibinfo {author} {\bibfnamefont {M.}~\bibnamefont
  {Pedersen}}, \bibinfo {author} {\bibfnamefont {S.~A.}\ \bibnamefont {{Van
  Langen}}}, \ and\ \bibinfo {author} {\bibfnamefont {M.}~\bibnamefont
  {B\"{u}ttiker}},\ }\href@noop {} {\bibfield  {journal} {\bibinfo  {journal}
  {Phys. Rev. B},\ }\textbf {\bibinfo {volume} {57}},\ \bibinfo {pages} {1838}
  (\bibinfo {year} {1998})}\BibitemShut {NoStop}%
\bibitem [{\citenamefont {Young}\ and\ \citenamefont
  {Clerk}(2010)}]{Young2010}%
  \BibitemOpen
  \bibfield  {author} {\bibinfo {author} {\bibfnamefont {C.~E.}\ \bibnamefont
  {Young}}\ and\ \bibinfo {author} {\bibfnamefont {A.~A.}\ \bibnamefont
  {Clerk}},\ }\href@noop {} {\bibfield  {journal} {\bibinfo  {journal} {Phys.
  Rev. Lett.},\ }\textbf {\bibinfo {volume} {104}},\ \bibinfo {pages} {186803}
  (\bibinfo {year} {2010})}\BibitemShut {NoStop}%
\bibitem [{\citenamefont {Ingold}\ and\ \citenamefont
  {Nazarov}(1992)}]{Ingold1992}%
  \BibitemOpen
  \bibfield  {author} {\bibinfo {author} {\bibfnamefont {G.}~\bibnamefont
  {Ingold}}\ and\ \bibinfo {author} {\bibfnamefont {Y.~V.}\ \bibnamefont
  {Nazarov}},\ }\bibfield  {title} {\enquote {\bibinfo {title} {{Charge
  tunneling rates in ultrasmall junctions}},}\ }in\ \href@noop {} {\emph
  {\bibinfo {booktitle} {Single Charge Tunneling}}},\ Vol.\ \bibinfo {volume}
  {294},\ \bibinfo {editor} {edited by\ \bibinfo {editor} {\bibfnamefont
  {H.}~\bibnamefont {Grabert}}\ and\ \bibinfo {editor} {\bibfnamefont {M.~H.}\
  \bibnamefont {Devoret}}}\ (\bibinfo  {publisher} {Plenum Press},\ \bibinfo
  {address} {New York},\ \bibinfo {year} {1992})\ Chap.~\bibinfo {chapter} {2},
  pp.\ \bibinfo {pages} {21--107}\BibitemShut {NoStop}%
\bibitem [{\citenamefont {Aguado}\ and\ \citenamefont
  {Kouwenhoven}(2000)}]{Aguado2000}%
  \BibitemOpen
  \bibfield  {author} {\bibinfo {author} {\bibfnamefont {R.}~\bibnamefont
  {Aguado}}\ and\ \bibinfo {author} {\bibfnamefont {L.}~\bibnamefont
  {Kouwenhoven}},\ }\href@noop {} {\bibfield  {journal} {\bibinfo  {journal}
  {Phys. Rev. Lett.},\ }\textbf {\bibinfo {volume} {84}},\ \bibinfo {pages}
  {1986} (\bibinfo {year} {2000})}\BibitemShut {NoStop}%
\end{thebibliography}

%

\end{document}